\documentclass[twocolumn,secnumarabic,amssymb,amsfonts,nobibnotes,aps,prl,showpacs]
{revtex4-1}
\usepackage{amsmath}
\usepackage{latexsym}
\usepackage {mathrsfs}
\usepackage{float}
\usepackage{amssymb}
\usepackage{amsfonts}
\usepackage{graphicx}
\usepackage{textcomp}
\usepackage{hyperref}
\usepackage {mathrsfs}
\usepackage {pifont}
 1
\begin{document}
\title{Finite-Size Effects in Dynamics: Critical vs Coarsening 
Phenomena}
\author{Subir K. Das$^\text{1,*}$, Sutapa Roy$^\text{1}$, 
Suman Majumder$^\text{1}$ and Shaista Ahmad$^\text{1,2}$}
\affiliation{$^1$Theoretical Sciences Unit, Jawaharlal Nehru 
Centre for Advanced Scientific Research, Jakkur P.O, Bangalore 
560064, India\\
$^2$ School of Physical Sciences, Jawaharlal Nehru University, 
New Delhi 110067, India}

\date{\today}

\begin{abstract} Finite-size effects in systems with diverging 
characteristic lengthscale have been addressed via state-of-the-art 
Monte Carlo and molecular dynamics simulations of various models 
exhibiting solid-solid, liquid-liquid and vapor-liquid transitions. 
Our simulations, combined with the appropriate application of 
finite-size scaling theory, confirm various non-trivial 
singularities in equilibrium dynamic critical phenomena and 
non-equilibrium domain coarsening phenomena, as predicted by 
analytical theories. We convincingly demonstrate that the 
finite-size effects in the domain growth problems, with 
conserved order parameter dynamics, is weak and universal, 
irrespective of the transport mechanism. This result is 
strikingly different from the corresponding effects in critical 
dynamics. In critical phenomena, difference in 
finite-size effects between statics and dynamics is also discussed.
\end{abstract}


\maketitle

\par
\hspace{0.2cm}In computer simulations, finite size of the systems poses 
enormous difficulty in studying problems where characteristic 
length scales diverge \cite{Fisher_crit,fisher2}. E.g., in 
equilibrium critical phenomena \cite{Onuki} the correlation 
length ($\xi$) diverges as
\begin {eqnarray}\label{1}
\xi \approx {\xi_0}\epsilon^{-\nu},
\end{eqnarray}
where $\epsilon=\lvert T-T_c \rvert/T_c$, $T_c$ being a critical 
temperature. On the other hand, when a homogeneous system is 
quenched inside the miscibility gap, the phase separation 
progresses via divergence of average domain-size, $\ell(t)$, as 
a function of time $(t)$ \cite{Onuki,Haasen,bray,Wadhawan} as
\begin {eqnarray}\label{2}
\ell(t) \approx At^\alpha,
\end{eqnarray}
where the exponent $\alpha$ depends upon the transport mechanism. 
This difficulty can, of course, be overcome via application of 
finite-size scaling theory \cite{Fisher_crit,fisher2,MC_book}. 
Nevertheless, it is of immense importance to learn the effects 
of finite system size, e.g., the study of nucleation and growth 
in nano-scopic systems, structure and dynamics in ultrathin films, 
etc., are of great independent interest. Also, an appropriate 
knowledge of the size effects helps judicial choice of the system 
size for the direct understanding of the problem in the 
thermodynamic limit so that any unexpected deviation from a 
prediction is not inappropriately attributed to the deficiency in 
system size.
\par
\hspace{0.2cm} While in static critical phenomena such problems 
are well addressed, the situation appears challenging in dynamics. 
It is certainly of fundamental importance to make a comparative 
study of finite-size effects in statics and dynamics. However, 
there are only a few computational studies 
\cite{Yethiraj,Chen,Das_prl,Horbach,Roy} of dynamic critical 
phenomena due to the fact that here, in addition to finite-size 
effects, the critical slowing down brings in another major hurdle. 
So finite-size effects are not appropriately probed and were 
thought to be same as in statics. On the other hand, despite a 
lot of simulation studies over several decades, the finite-size 
scaling theory in non-equilibrium domain coarsening problems found 
only rare application \cite{Heerman,Majumder1,Majumder2} and the 
finite-size effects in this type of problems remained a challenging 
issue.
\par
\hspace{0.2cm}In this letter, in addition to confirming results 
for various singular behaviors in critical and coarsening phenomena, 
we address the issue of finite-size effects in these two sets of 
problems under a very general framework. For the critical phenomena, 
we present results for both static and dynamic properties from 
Monte Carlo (MC) and molecular dynamics (MD) simulations of a 
liquid-liquid (LL) phase transition in a binary Lennard-Jones (LJ) 
system \cite{Horbach}. On the non-equilibrium front, results for 
kinetics of phase separation are presented for LL transition using 
the same LJ system, for vapor-liquid (VL) phase transition in a single 
component LJ model \cite{Majumder3} as well as for a solid binary mixture 
(SS) using Ising model. An unified picture for the finite-size effects 
is obtained in this case despite different transport mechanisms leading to 
different values \cite{Onuki,Haasen,bray} of $\alpha$ in solids and 
fluids. Finally, contrasting observation between the finite-size 
effects in the non-equilibrium dynamics and the equilibrium critical 
dynamics is discussed. While these interesting results are expected 
to initiate further theoretical studies, the methods used here will 
be of significant importance in fluid dynamics, glass transition and 
other condensed matter systems.
\par
\hspace{0.2cm}For the binary $(A+B)$ liquid, we consider a model 
(belonging to the $3-$d Ising critical universality class, d being 
the space dimension) where particles at continuous positions 
${\vec r}_i$ and ${\vec r}_j$, in a periodic box of linear dimension 
$L$ (in units of the LJ particle diameter), interact (for $r<r_c$) via
\begin {eqnarray}\label{pot}
u(r=|{{\vec r}_i}-{{\vec r}_j}|)=U(r)-U(r_c)-(r-r_c)
{\frac {dU}{dr}}|_{{r}=r_c},~~~~
\end{eqnarray}
where 
\begin {eqnarray}\label{pot1}
U(r)=4J_{\alpha\beta}[(d_{\alpha\beta}/{r_{ij}})^{12}-
(d_{\alpha\beta}/{r_{ij}})^6];
~~\alpha,\beta \in A,B
\end{eqnarray}
is the standard LJ potential with the pairwise interaction 
strength $J_{AA}=J_{BB}=2J_{AB}=J$ and LJ diameter 
$d_{AA}=d_{BB}=d_{AB}=d_0$. The second term on the right hand 
side of Eq.(\ref{pot}) stands for a truncation and shifting 
of the potential at $r=r_c$ which we chose to be $2.5d_0$. 
Finally, the third term ensures that both the potential and the 
force are continuous all along. We simulate this completely 
symmetric model at a high density $\rho={{N{d_0}^3}/{V}}=1$ 
that exhibits a liquid-liquid transition at the critical 
parameters ${k_B}{T_c}\simeq1.423J$ and 
$x_A=x_A^c=1/2$ $(x_A=N_A/N;~N=N_A+N_B,~ N_\alpha~
\textrm{being the number of particles of species}~\alpha)$. 
For the sake of convenience, in the rest of the paper we set 
$k_B$, $J$, $d_0$ as well as the equal mass $(m)$ of the 
particles to unity which in turn sets the LJ time unit 
$t_0=(md_0^2/{J})^{1/2}=1$. Phase behavior and static 
concentration susceptibility $(\chi)$ for 
this model were calculated from MC simulation in the semigrand 
canonical ensemble \cite{MC_book,Horbach} whereas the transport 
properties were obtained from relevant Green-Kubo relations 
\cite{Hansen,bray} by using outputs from MD simulations 
(for $T>T_c$) in microcanonical ensemble that perfectly 
preserves hydrodynamics. For the VL transition 
$(T_c \simeq 1, \rho_c\simeq0.32)$ \cite{Hansen1}, the same 
model with only one species of particles was used. To probe 
the hydrodynamic effects in the kinetics of fluid phase 
separation, both the LL and VL systems were studied via 
MD simulations in NVT ensemble by quenching a homogeneous 
system (of critical composition or density) below the 
respective critical temperatures. Here the temperature was 
controlled via Nos\'{e}-Hoover thermostat \cite{Frenkel} 
that is known to preserve hydrodynamics well. Finally, for 
the SS case, spin-$\frac{1}{2}$ critical ($50:50$) Ising model 
with hamiltonian
\begin {eqnarray}\label{hamil}
H=-J\sum\limits_{<ij>}{S_i}{S_j};~S_i=\pm1,~J=1,
\end{eqnarray}
was studied via Kawasaki-exchange MC \cite{MC_book} simulation, 
where exchange between two neighboring spins consisted a 
trial move that preserves composition of up (A) and down (B) 
spins (particles). This introduces a diffusive dynamics which 
is responsible for domain coarsening in solid mixtures. An MC 
step (MCS) consisted of $L^d$ trial moves.
\begin{figure}[htb]
\centering
\includegraphics*[width=0.36\textwidth]{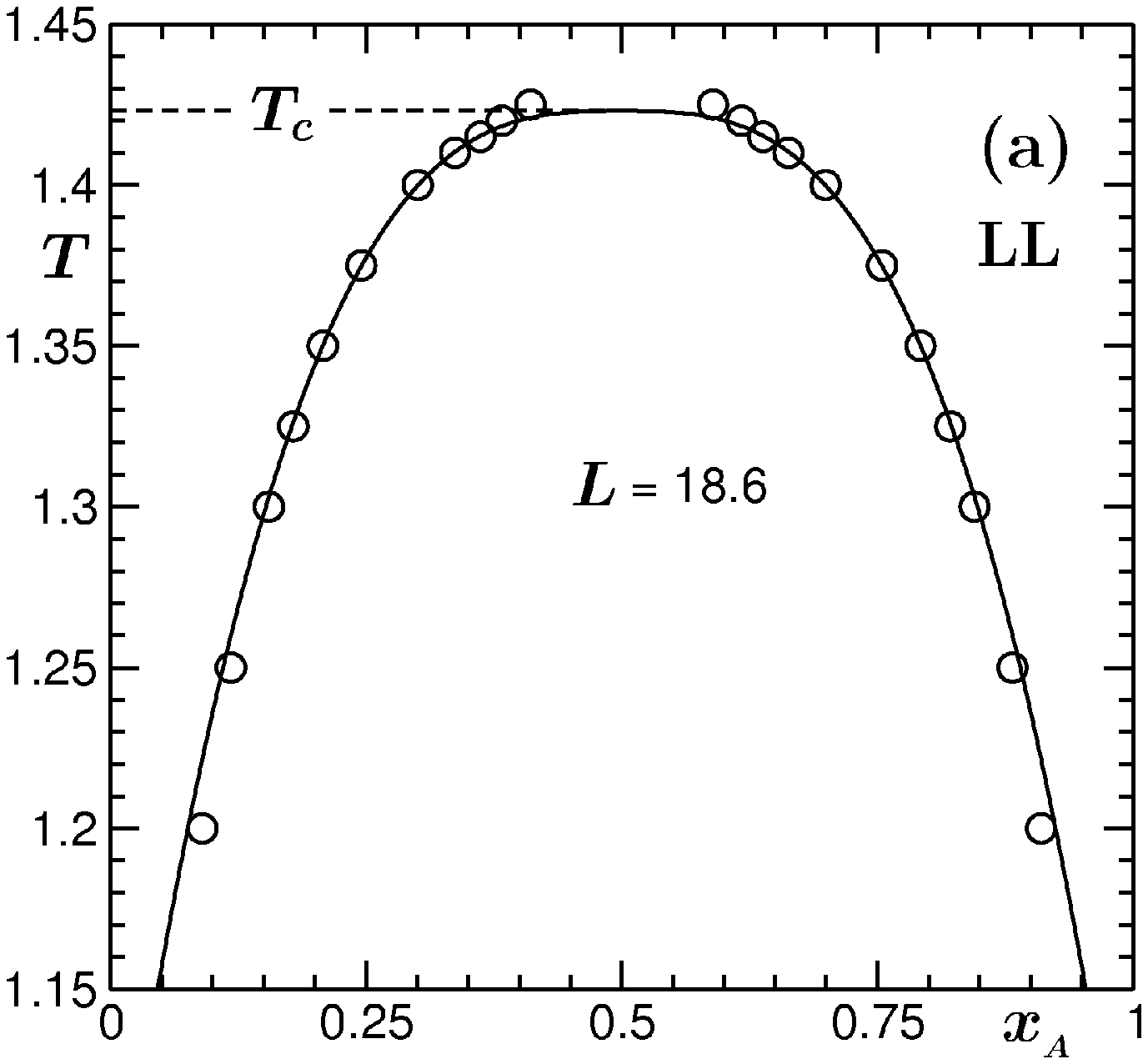}
\vskip 0.3cm
\includegraphics*[width=0.36\textwidth]{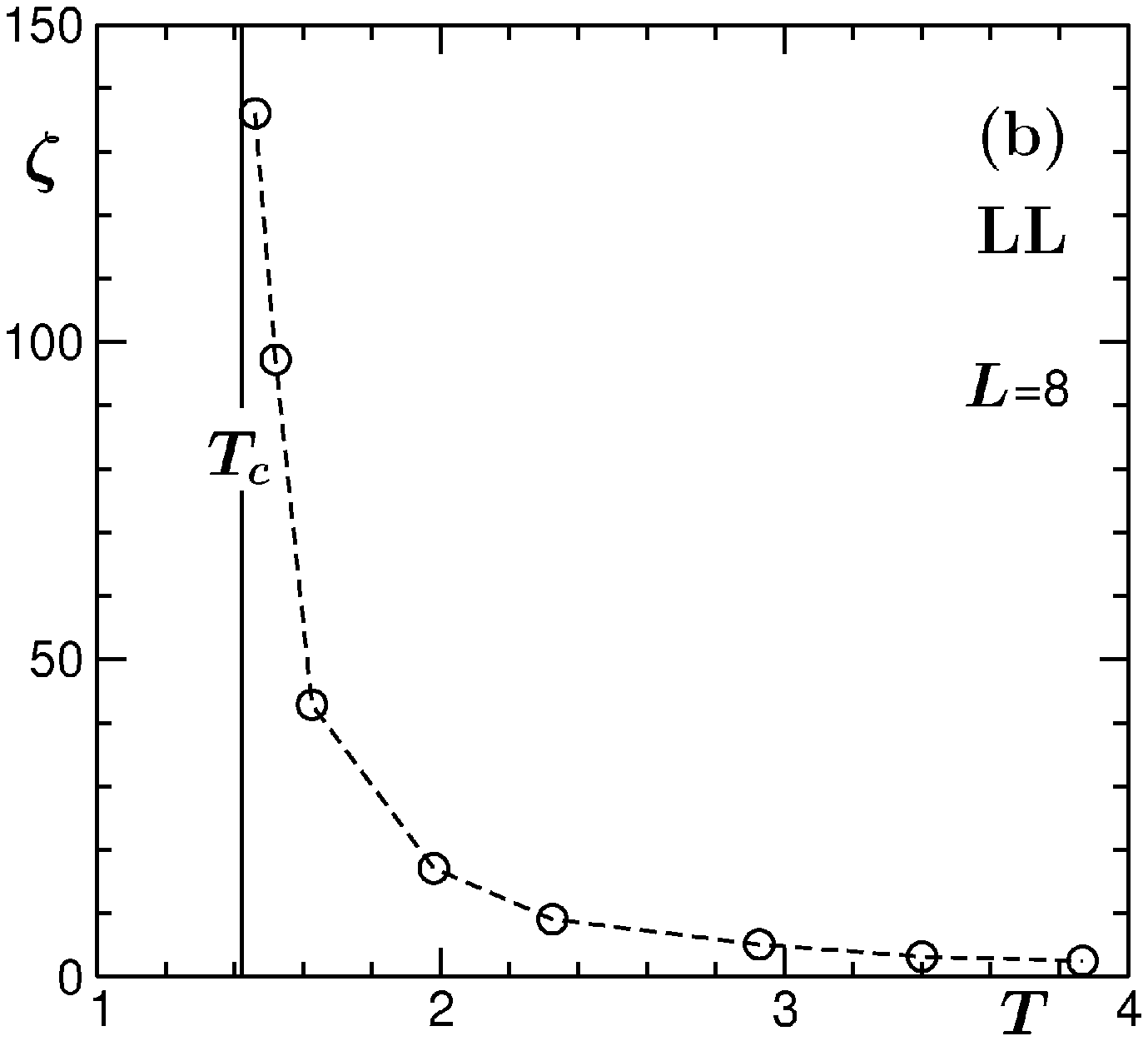}
\caption{\label{fig1}(a) Phase behavior of the binary fluid 
model in temperature-concentration plane. The solid line there 
is a fit to the form $m=|{\frac{1}{2}}-x|\sim \epsilon^\beta;~
\beta=0.325$ (Ising value), by choosing a region close to the 
critical point but unaffected by finite-size of the system. 
This provides $T_c=1.423\pm0.002$. (b) Plot of bulk viscosity 
$(\zeta)$ vs temperature, for the same model, at the critical 
composition, for a system size $L=8$. The vertical line 
corresponds to the critical temperature.}
\end{figure}
\par
\hspace{0.2cm}The primary quantity of interest in the 
non-equilibrium phenomena is the average domain size, $\ell(t)$, 
which, unless otherwise mentioned, was obtained from the first 
zero crossing of the two-point equal-time correlation function 
\cite{Wadhawan} $C(r,t)$ where $r$ is the distance between two 
points. For phase ordering dynamics with conserved order-parameter, 
while $\alpha=1/3$ for diffusive domain-coarsening (which is the only 
coarsening mechanism in binary solids) \cite{Haasen,bray,Wadhawan}, 
for fluids \cite{Majumder3,Siggia,furukawa,Shaista} hydrodynamic 
mechanism plays important role and there a diffusive growth is 
followed by a viscous hydrodynamic one with $\alpha=1$ and then 
by an inertial hydrodynamic regime with $\alpha={2}/{3}$. 
On the other hand, for critical dynamics we present results 
for the bulk viscosity $(\zeta)$ (we observe similar size effects 
in other relevant transport properties as well, however, do not 
present those here for the sake of brevity). Note that, analogous 
to Eq. (\ref{1}), the critical singularity of any thermodynamic 
or dynamic property, $X$, is quantified as  
\begin {eqnarray}\label{xeq}
X \approx {X_0}\epsilon^{-x} 
\end{eqnarray}
where $x=0.63,~1.239,~\textrm{and}~1.82$ for $\xi$, $\chi$ and 
$\zeta$, respectively \cite{Onuki,Das_prl,Roy,Onuki_pre,Olchowy,
Luettmer,Hao,Iwanowsaki,Mizzaev}, for 3-d Ising universality class. 
Unless otherwise mentioned, all results are presented in this 
space dimension.
\begin{figure}[htb]
\centering
\includegraphics*[width=0.36\textwidth]{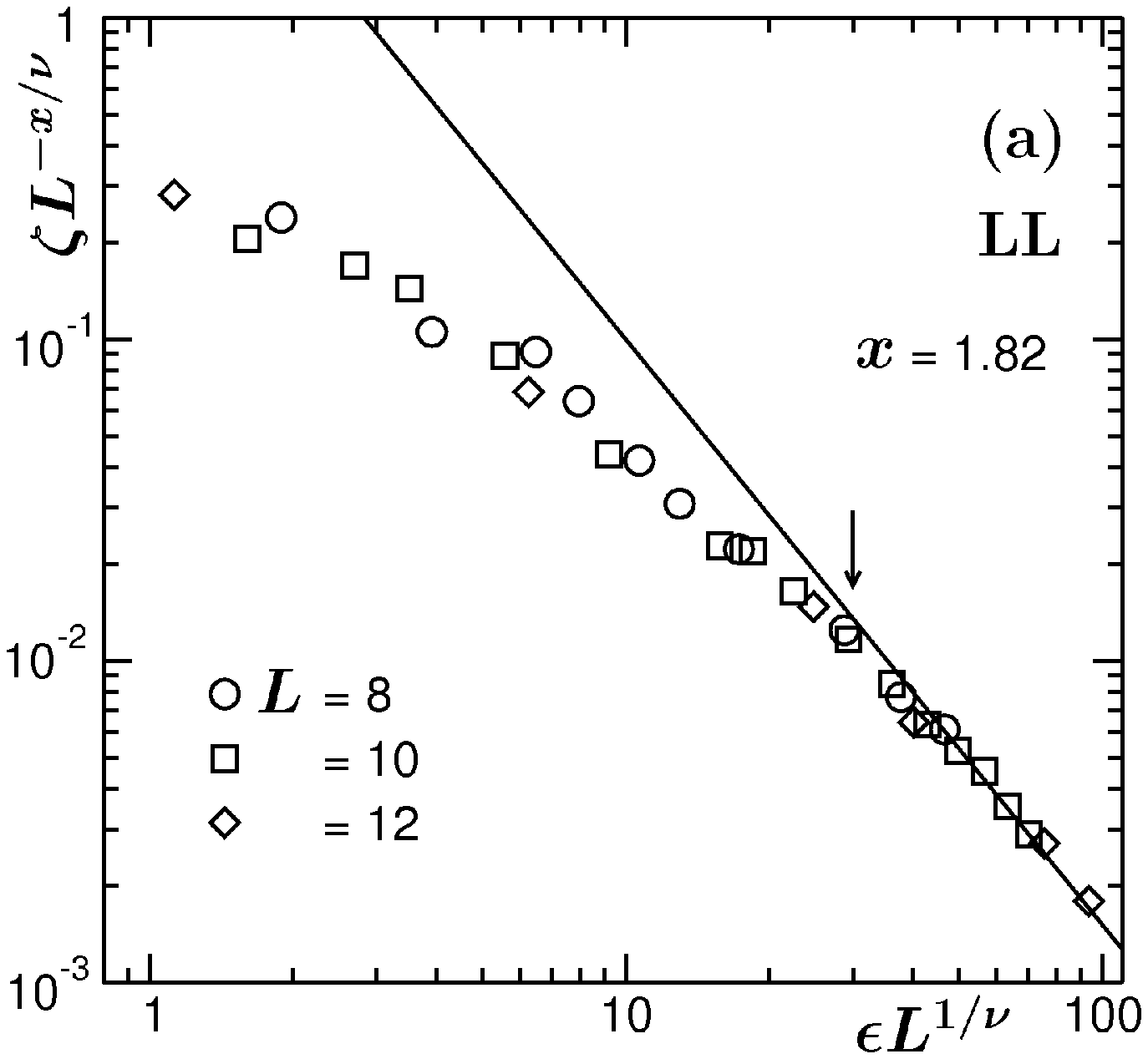}
\vskip 0.3cm
\includegraphics*[width=0.36\textwidth]{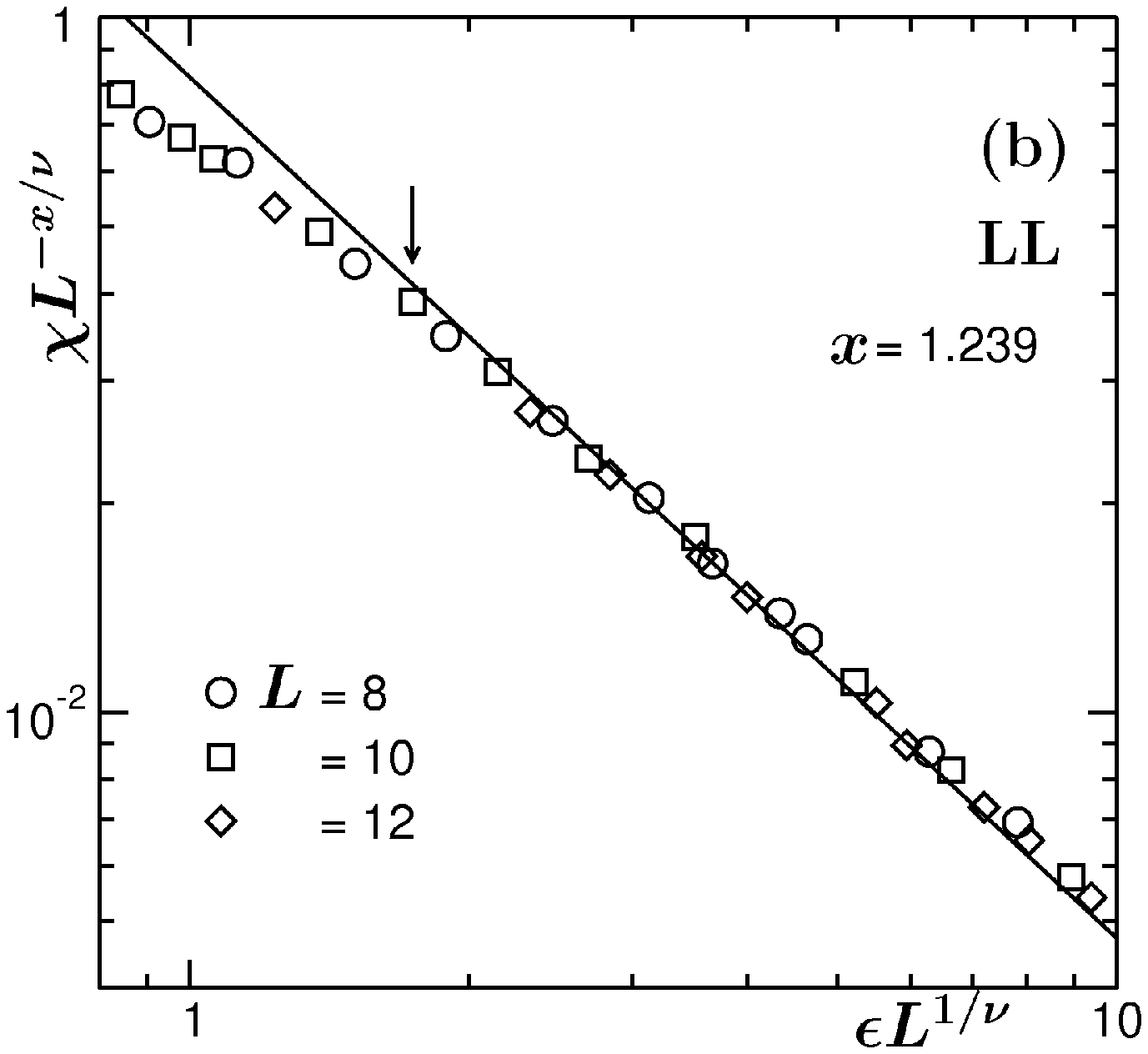}
\caption{\label{fig2}(a) Finite-size scaling plot of $\zeta$. 
Here ${\zeta}L^{{-1.82}/{\nu}}$ is plotted vs $\epsilon L^{1/\nu}$ 
by taking data from three different system sizes. The solid line 
there has a power-law behavior with exponent $-1.82$. (b) Same as 
(a) but for $\chi$ by using $x=1.239$.}
\end{figure}
\par
\hspace{0.2cm}For the sake of completeness, the phase diagram 
is shown in Fig.\ref{fig1}(a), for a system of size $L=18.6$. 
The continuous line there is a fit to the form 
$m=|\frac{1}{2}-x| \sim \epsilon^{\beta}$ by fixing $\beta$ to 
its $3-$d Ising value $0.325$. Note that data only close to the 
critical point but unaffected by finite-size of the system were 
used. From this exercise, we obtain $T_c=1.423\pm 0.002$. 
In Fig.\ref{fig1}(b) we present the simulation results for the 
bulk viscosity $(\zeta)$ as a function of temperature, for the 
binary LJ system, with $L=8$.  The data for $\zeta$ shows a sharp 
increase close to the critical point. While this signifies a 
strong critical divergence, a least-square fitting using the form 
(\ref{xeq}) gives an exponent much smaller than $1.82$ which could 
well be due to strong finite-size effects. Thus, an appropriate 
understanding of the result calls for the following 
finite-size scaling analysis. (Note that a general 
discussion of finite-size scaling analysis in this context is 
provided in Ref. \cite{Chimowitz}. However, for the sake of 
completeness we briefly discuss it here.) At the critical point, 
the singularity of $X$, as a function of the system size, is 
characterized as
\begin {eqnarray}\label{3}
X=A_{0}L^{x/\nu},
\end{eqnarray}
where $A_{0}=X_{0}/(2\xi_{0})^{x/\nu}$ and we have used the 
fact that $\xi=L/2$ at $T_c$. Away from $T_c$, one needs to 
introduce a scaling function $Y(y)$ to write
\begin {eqnarray}\label{4}
X=L^{x/\nu}Y(y),
\end{eqnarray}
where $y$ is a function of the dimensionless variable ${{L}/{\xi}}$.
While, $Y(y)=A_{0}$ at $T=T_c$, for the convenient choice 
$y=\left({{L}/{\xi}}\right)^{1/\nu}(\propto {\epsilon L^{1/\nu}})$ 
and $T \gg T_{c}~(L \gg \xi)$, one must have
\begin {eqnarray}\label{5}
Y(y) \sim y^{-x},
\end{eqnarray}
so that Eq.(\ref{xeq}) is recovered. Thus, when $XL^{-x/\nu}$ is 
plotted vs $y$, in addition to collapse of data coming from 
different system sizes, one should obtain a power-law behavior 
with the exponent $-x$, for $y \gg 0$. A deviation from this 
power-law, for smaller $y$, signals the onset of finite-size 
effects. In Fig.\ref{fig2} (a) and (b) this is demonstrated for 
$\zeta$ and $\chi$, for the binary LJ model. Here we have fixed 
$x$ at the respective values and used $T_c$ as an adjustable 
parameter. The best collapse of data is obtained at 
$T_c \simeq 1.421$ which, of course, is consistent with the 
value quoted in the caption of Fig.{\ref{fig1}. For large enough 
$y$, both the quantities are consistent with the expected critical 
behaviors represented by the solid lines.  However, the striking 
difference between the onsets of significant finite-size effects 
(marked by arrows), quantified by the ratio $z_0=\xi/L$, should be 
noted. It appears that $z_0={\mathcal O}(1)$ for $\chi$ and 
$={\mathcal O}(10^{-1})$ for $\zeta$.
\begin{figure}[htb]
\centering
\includegraphics*[width=0.33\textwidth]{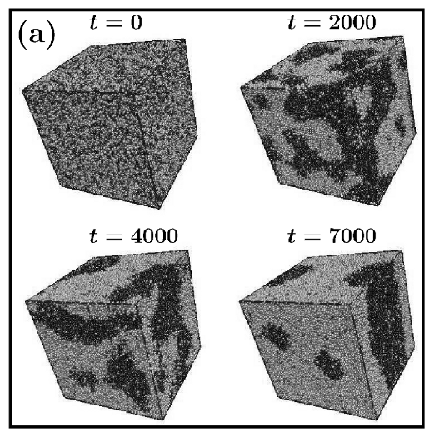}
\vskip 0.3cm
\includegraphics*[width=0.36\textwidth]{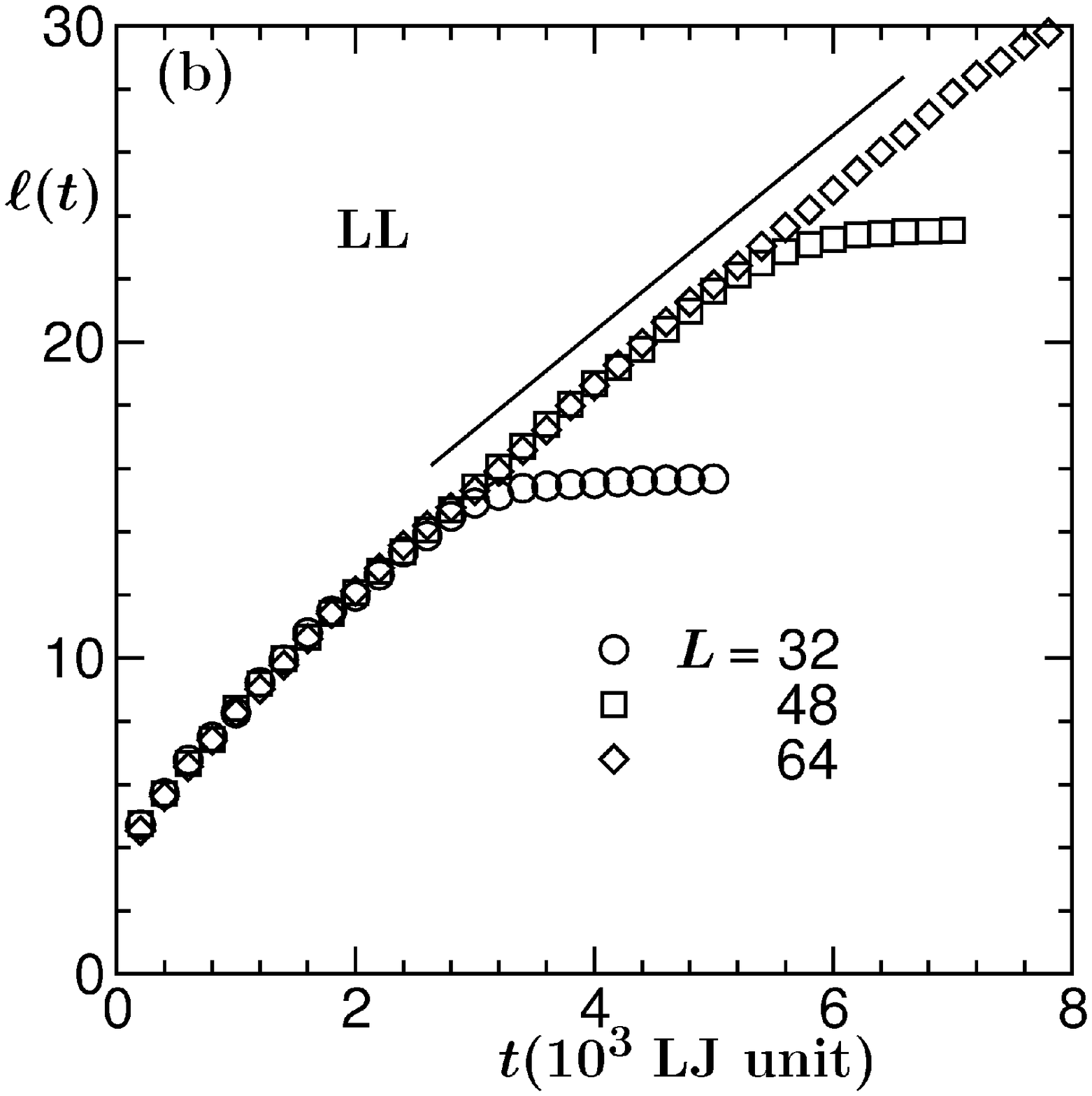}
\caption{\label{fig3}(a) Evolution snapshots from four different 
times after quenching a homogeneously mixed $50:50$ binary fluid, 
confined in a box of linear dimension $L=64$, into the miscibility gap. 
(b) Plot of $\ell(t)$ as a function of $t$, for the model of (a). 
Results from $3$ different system sizes are displayed. The solid 
straight line there corresponds to the linear viscous hydrodynamic 
growth. All results correspond to a temperature $T=0.77T_c$.}
\end{figure}
\par
\hspace{0.2cm}We start the discussion of non-equilibrium phenomena 
from Fig.\ref{fig3}(a) which demonstrates the formation and growth 
of $A$-rich and $B$-rich domains for $L=64$ in a binary LJ fluid. 
Note that the thermal noise seen in the snapshots create difficulty 
in accurate estimation of the average domain size. This problem was 
avoided via following procedure. First we have mapped the continuum 
system into a $L^3$ lattice where a site occupied by an $A$-particle 
was assigned a spin value $+1$, otherwise $-1$. Further, a majority 
spin rule, where the value of the spin at a site was replaced by the 
sign of the majority of the spins around it, was used to eliminate 
the noise. Note that for LL, VL, as well as SS cases, the quantitative 
analysis was done by using the noise-free, so called ``pure domain", 
snapshots. 
\par
\hspace{0.2cm}In Fig.\ref{fig3}(b) we show the plots of $\ell(t)$ 
(in units of LJ diameter) vs $t$, for three different system sizes, 
obtained after quenching a homogeneously mixed system to a temperature 
$T=0.77T_c$. For $L=64$, it is clearly seen that $\ell(t)$, 
beyond $t=2\times10^3$, is growing quite linearly, consistent with an 
expected viscous hydrodynamic behavior. Even though, in this case the 
finite-size effects did not appear yet, for $L=32$ and $48$, the flat 
natures of the data towards the ends indicate that the equilibriums 
have been reached. It is interesting to notice that the results for 
the smaller systems follow the larger ones almost till the saturation 
limit. This is already suggestive of only weak size effects. 
Nevertheless, to correctly quantify it, we again take the help of 
finite-size scaling theory which, in addition, will provide further 
concrete information about the growth law.
\par
\hspace{0.2cm}In the present case, a finite-size scaling tool could 
be constructed by making the obvious identification of $1/t$ with 
$\epsilon$ and $\ell(t)$ with $\xi$. Then the relation equivalent to 
Eq. ({\ref{4}}) is
\begin {eqnarray}\label{6}
\ell(t)=LY(y);~~y&=&\left({{L}/{\ell}}\right)^{1/\alpha}
\propto{{L^{1/\alpha}}/{t}}.
\end{eqnarray}
Thus, when $\ell(t)L^{-1}$ is plotted vs $y$, for large $y$, a 
power-law behavior with an exponent $-\alpha$ should be obtained. 
For the sake of convenience, we demonstrate this first for the 
$2-d$ Ising model, quenched to the temperature $T=0.6T_c$.
\par
\hspace{0.2cm}In Fig. \ref{fig4}(a), we show the direct plots of 
$\ell(t)$ (in units of lattice constant) vs $t$, for two different 
system sizes. In Fig. \ref{fig4}(b), we show a trial plot for the 
scaling behavior contained in Eq. (\ref{6}) where, of course, 
we have correctly substituted $L$ by the corresponding maximum 
domain length [see Fig. \ref{fig4}(a)], $\ell_{\textrm{max}}$, 
that represents the equilibrium limit. 
\begin{figure}[htb]
\centering
\includegraphics*[width=0.36\textwidth]{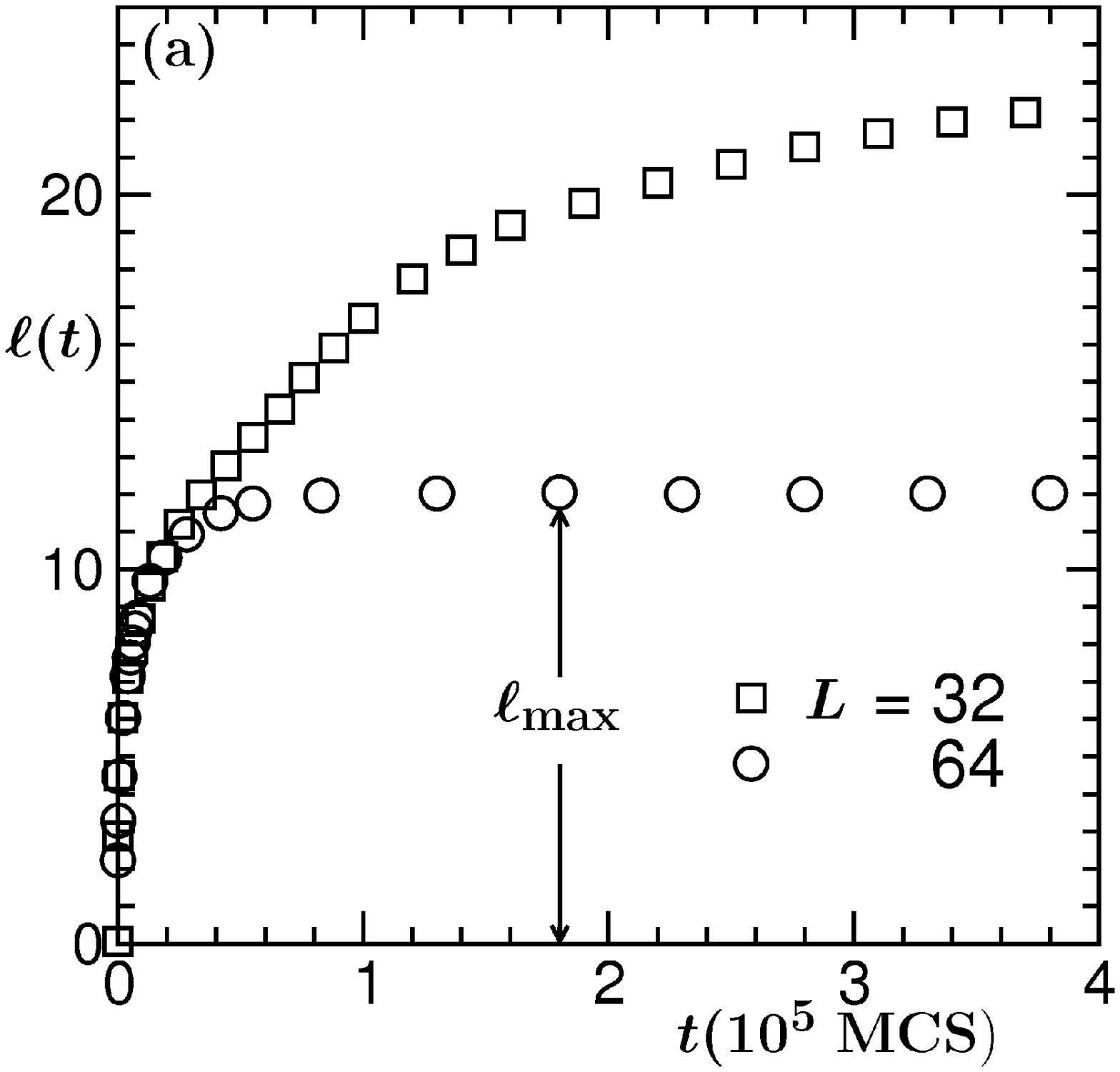}
\vskip 0.3cm
\includegraphics*[width=0.36\textwidth]{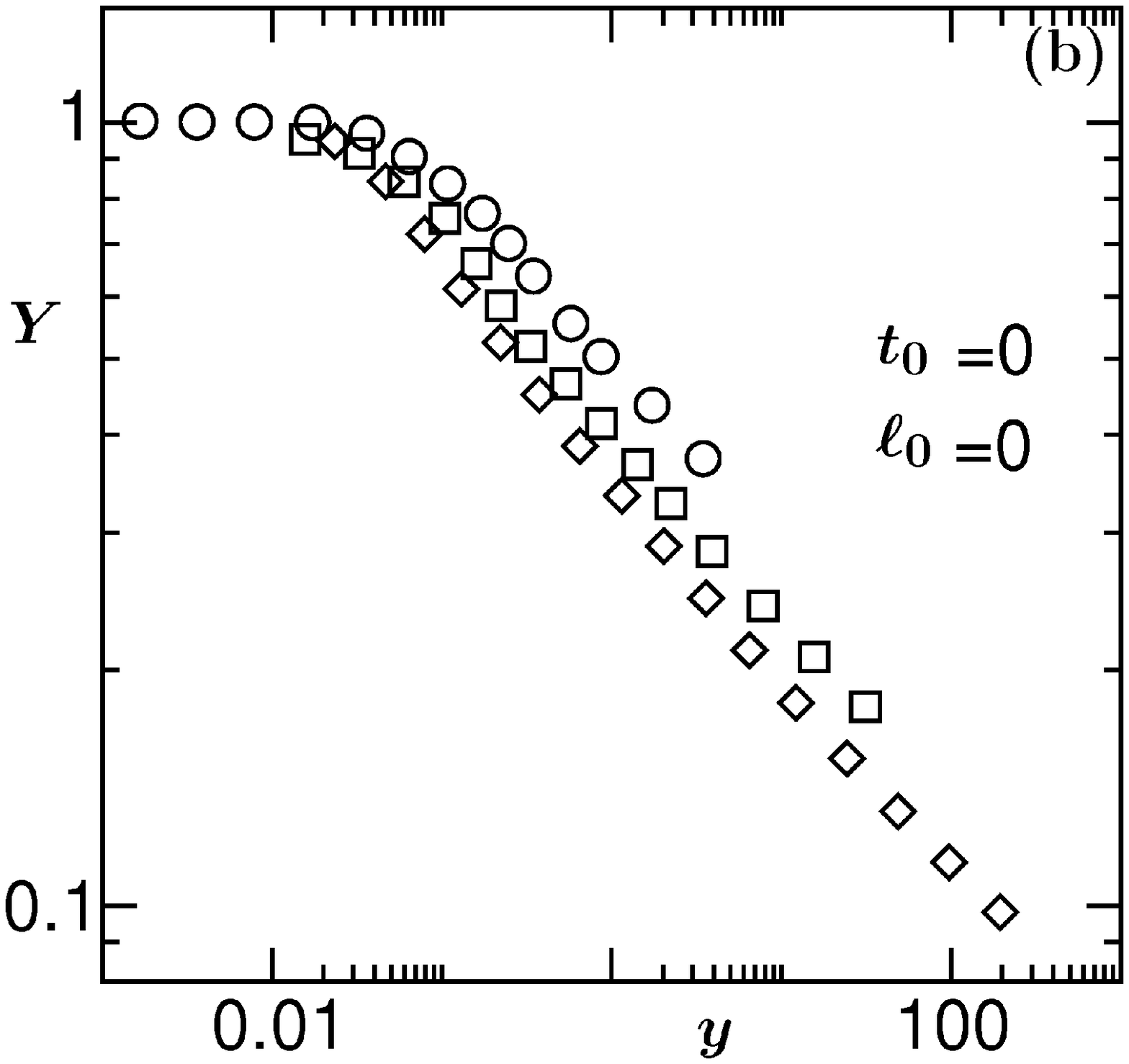}
\vskip 0.3cm
\includegraphics*[width=0.36\textwidth]{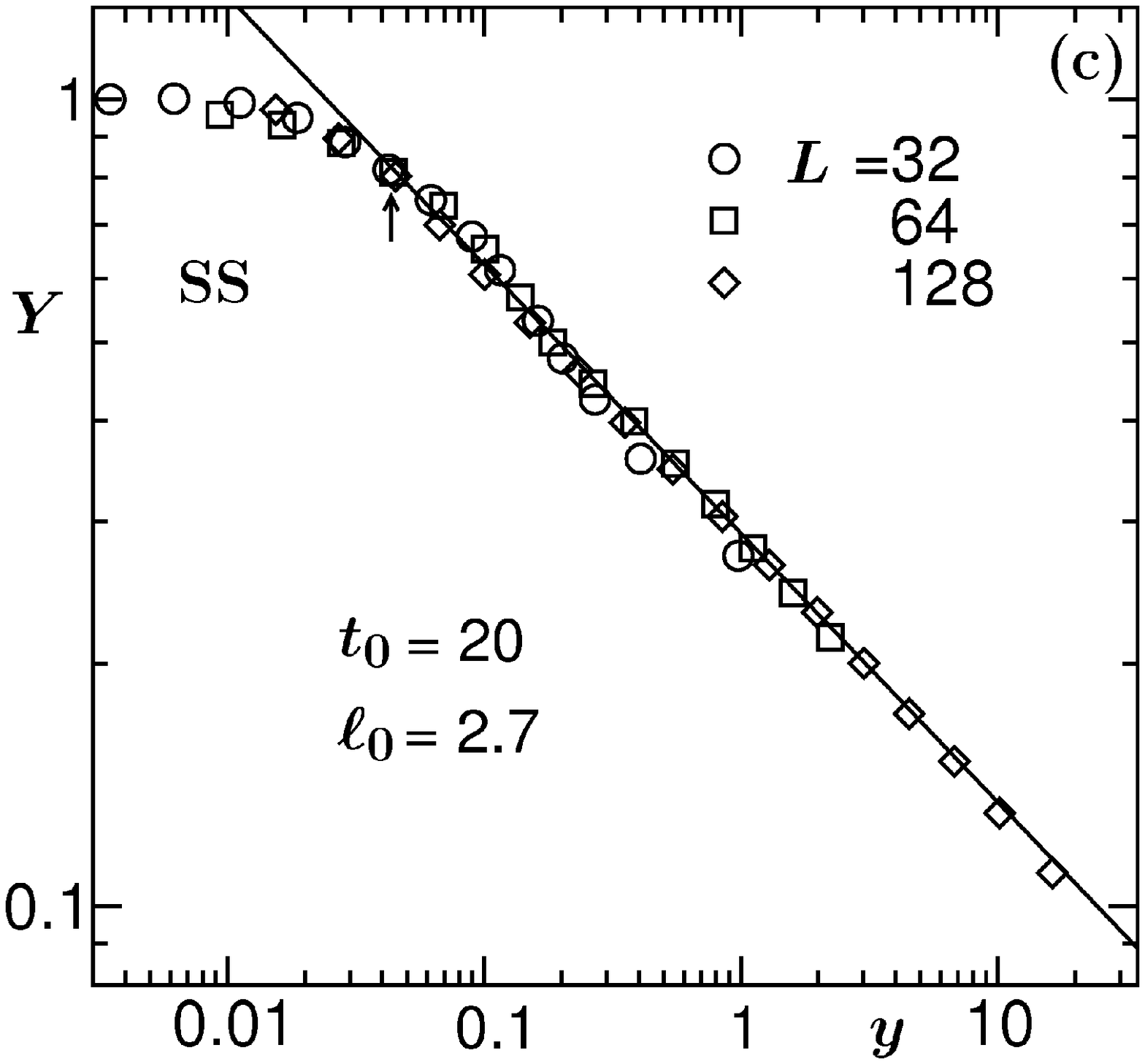}
\caption{\label{fig4} (a) Plot of length scale $\ell(t)$ as a 
function of $t$, for $2-$d Ising model, for two different system 
sizes as shown in the figure. (b) Finite-size scaling plot of 
$\ell(t)$, in accordance with Eq. (\ref{6}) using $3$ different 
system sizes with $t_0$ set to zero. (c) Same as (b) but $t_0$ 
(thus $\ell_0$) was varied to obtain the optimum data collapse.}
\end{figure}
We have fixed 
$\alpha={1}/{3}$, as expected for diffusive domain growth in 
Kawasaki-exchange Ising model. Here, very poor quality of data 
collapse for large values of $y$ is due to the fact that the 
systems do not enter the scaling regime immediately. In fact, 
after the quench the system requires a while to become unstable 
to fluctuations. Of course, this non-overlapping behavior will 
not be seen if one has data over many decades in time for a 
significantly large system. But this will be prominent for small 
systems. Thus, for a correct analysis, one needs to subtract a 
time $t_0$ from $t$ (and corresponding length $\ell_0$ from 
$\ell_{\textrm{max}}$ as well as $\ell(t)$) to work with only 
the scaling part. Note that $\ell_0$ is independent of time and 
is analogous to a weakly temperature dependent background 
contribution in critical phenomena. The correct value of $t_0$ 
(and so $\ell_0$) must correspond to the optimum data collapse. 
This is illustrated in the Fig. \ref{fig4}(c) where excellent 
collapse is obtained for $t_0=20$. The solid line there has the 
form $y^{-\alpha}$ $(\alpha=1/3)$ with which, for $y \gg 0$, 
the simulation results are perfectly consistent. The point of 
deviation of the data from this solid line provides us with 
the information about the onset of finite-size effects at 
$\ell(t) \simeq 0.77\ell_{\textrm{max}}$ which is informative 
of much weaker size effects compared to previous understanding. 
This now needs to be seen if this small finite-size effect is 
characteristic of only the simple Ising model or of more general 
validity. To investigate this, in the following we look back at 
the domain coarsening in fluids.

\begin{figure}[htb]
\centering
\includegraphics*[width=0.36\textwidth]{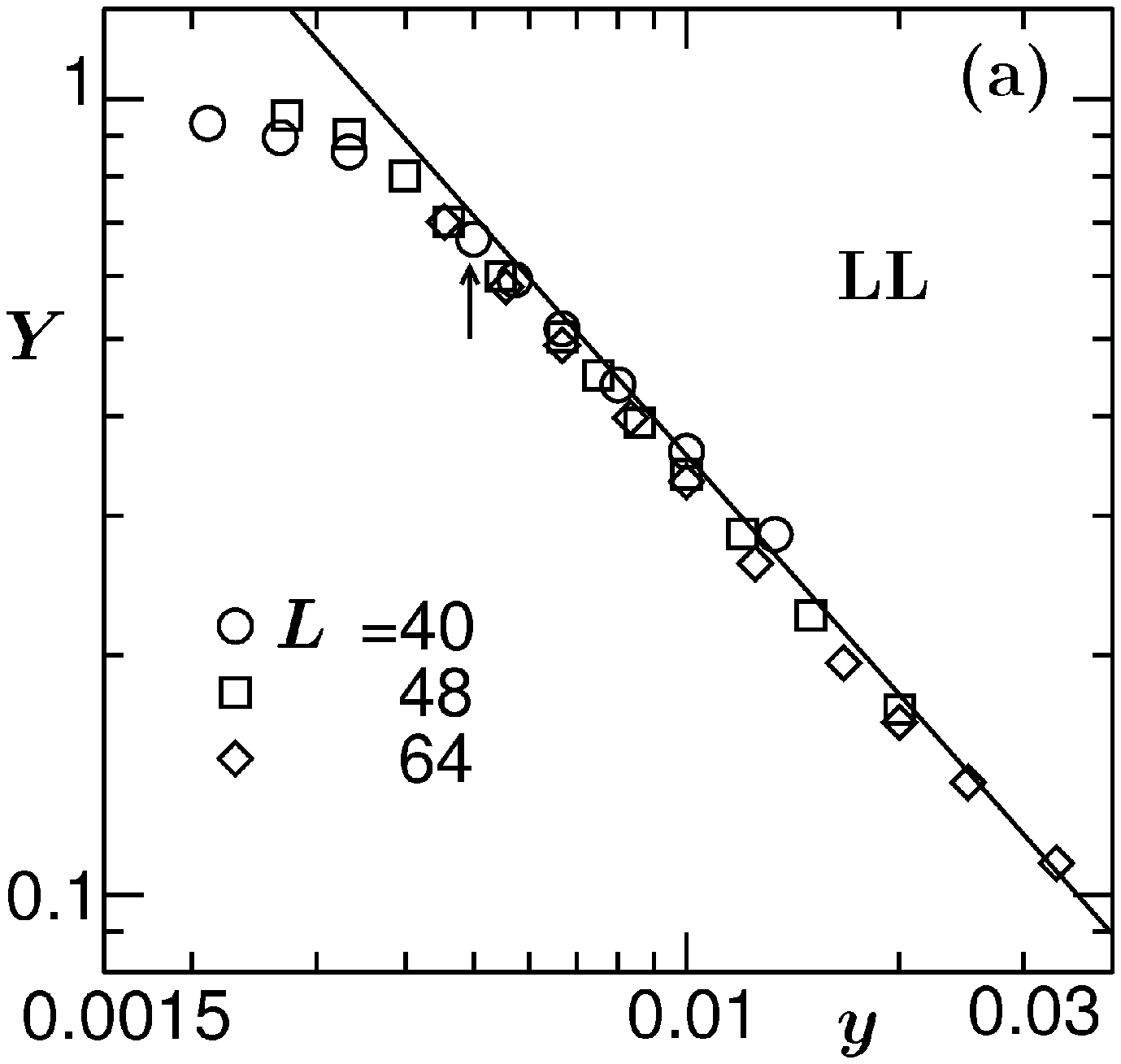}
\vskip 0.3cm
\includegraphics*[width=0.36\textwidth]{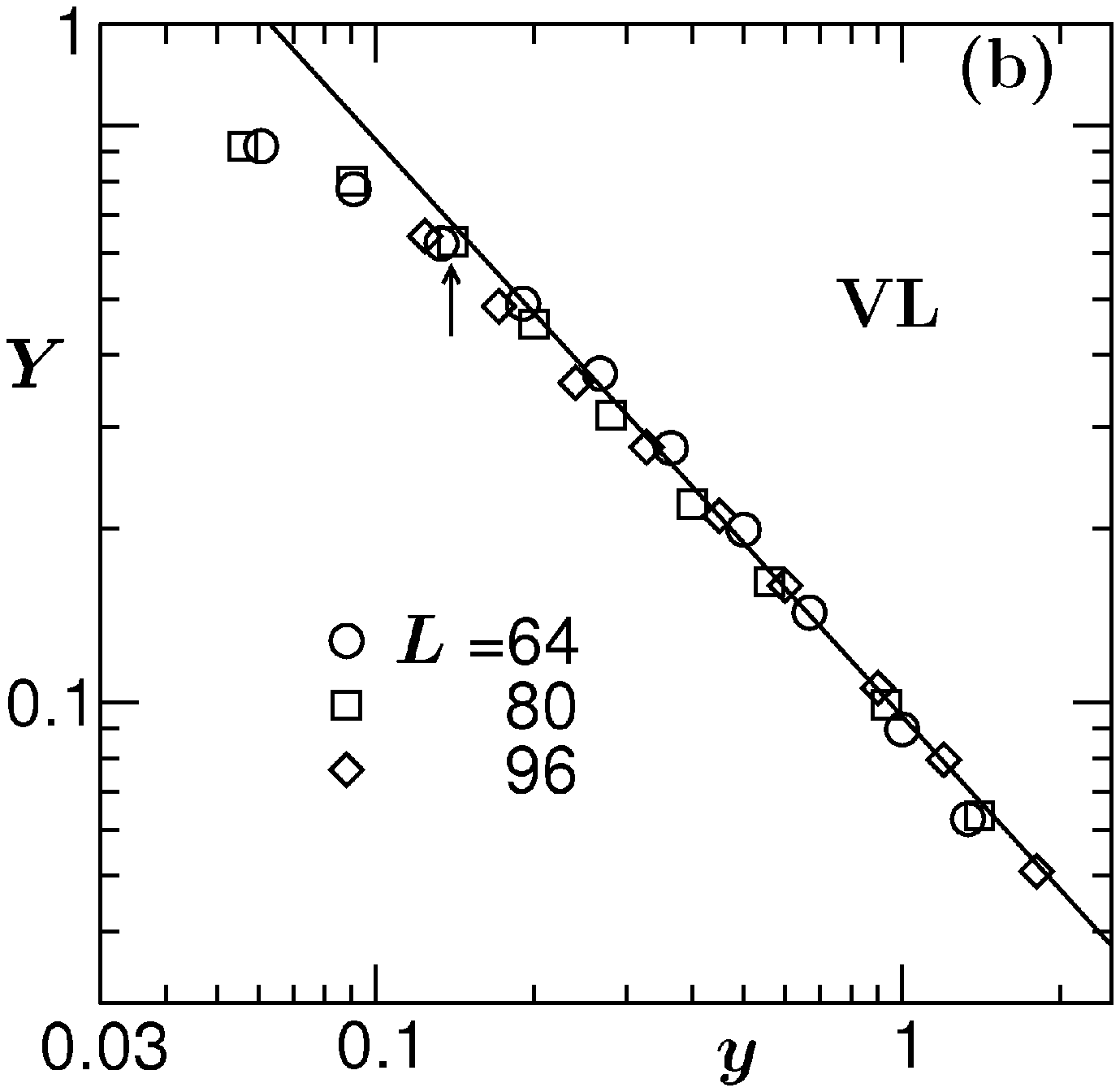}
\caption{\label{fig5}Finite-size scaling plots of length scale 
data for the (a) binary LJ model and (b) single component 
LJ model. In (b), $\ell(t)$ was obtained from the first moment 
of the domain-size distribution function.}
\end{figure}
\par
\hspace{0.2cm}In Fig. \ref{fig5} (a), we present the scaling 
plot of $\ell(t)$ for the binary LJ system of Fig. \ref{fig3}. As 
discussed, in fluid phase separation a diffusive domain growth is 
followed by a linear viscous growth and further by an inertial 
regime with an exponent $\alpha={2}/{3}$. Due to the obvious 
difficulty in dealing with significantly large system size for a 
long time, we are unable to observe the inertial growth. On the 
other hand, a gradual crossover to the linear regime from very 
early time does not allow us to observe a pure diffusive domain 
growth. Thus the focus in this exercise is to obtain a concrete 
answer for the linear behavior. On this occasion, a perfect data 
collapse could be obtained when the correct length ($\ell_c$) and 
time ($t_c$), corresponding to crossover from diffusive to viscous 
regime, are subtracted. In Fig. \ref{fig5} (a), the best collapse 
is obtained (by fixing $\alpha=1$) for $t_{c}=2 \times 10^3$ 
($\ell_c \simeq 12$) which could also be appreciated from 
Fig. \ref{fig3}. The consistency of the master curve with the solid 
line $(y^{-1})$ provides further confirmation about the linear 
behavior. A deviation from this solid line occurs at 
$\ell(t) \simeq 0.8 \ell_{\textrm{max}}$ which is very similar to 
the Ising model (SS) scenario. In Fig. \ref{fig5} (b), analogous 
exercise is demonstrated for the VL transition in a single component 
LJ fluid, for which we obtain $t_c=50$. Here also, the presence of a 
linear viscous hydrodynamic growth is confirmed and we quantify the 
appearance of finite-size effects at 
$\ell(t) \simeq 0.78 \ell_{\textrm{max}}$. Needless to say, the 
scaling would have failed if we were able to reach the inertial 
hydrodynamic regime by running a much bigger system for significantly 
longer period of time.
\par
\hspace{0.2cm}In summary, we have presented comprehensive results 
for the structure and dynamics from the state-of-the-art computer 
simulations of models exhibiting various phase transitions. It is 
demonstrated that the finite-size effects in equilibrium critical 
dynamics is much stronger compared to thermodynamics. Even though we 
believe that the finite-size effects in critical dynamics should be 
universal, as in the static case, this needs to be checked. On the 
other hand, the results for the non-equilibrium coarsening phenomena 
is suggestive of only weak size effects which is in sharp contrast 
with traditional understanding. Also, this non-equilibrium size 
effect appears to be universal for coarsening in solid-solid, 
liquid-liquid and vapor-liquid phase transitions which we have 
quantified to be appearing at 
$\ell(t) \simeq (0.8 \pm 0.1)\ell_{\textrm{max}}$. This result should 
be compared with the work of Ref. \cite{Heerman} where it was pointed 
out that the finite-size effect is strong. Note that in the latter 
work, only an off-critical composition was used and a final conclusion, 
for the appearance of finite-size effects, was drawn as a fraction of 
the system size $L$. Thus, the problem of off-critical composition may 
be revisited for more accurate conclusions in terms of 
$\ell_{\textrm{max}}$. Also, it would be interesting to test our claim 
about the universally weak finite-size effects via further studies at 
different quench depth. 
\par
\hspace{0.2cm}Further, via appropriate applications of finite-size 
scaling theory, we show that our results are consistent with the expected 
theoretical predictions related to the divergence of relevant quantities, 
both for equilibrium and non-equilibrium dynamics, in the thermodynamic limit.
\par
\acknowledgments 
\hspace{0.2cm}The authors acknowledge grant No. SR/S2/RJN-$13/2009$ of 
the Department of Science and Technology, India, for financial support. 
SR and SM acknowledge Council of Scientific and Industrial Research 
and SA acknowledges University Grants Commission for financial support. 
SA is grateful to JNCASR for hospitality during her research visits. 
\newline
$*$~das@jncasr.ac.in

\end{document}